# Magnon driven skyrmion dynamics in antiferromagnets: The effect of magnon polarization

Z. Jin[1], C. Y. Meng[1], T. T. Liu[1], D. Y. Chen[1], Z. Fan[1], M. Zeng[1], X. B. Lu[1], X. S. Gao[1], M. H. Qin[1,3*], and J. –M. Liu[1,2]

[1]*Institute for Advanced Materials, and Guangdong Provincial Key Laboratory of Optical Information Materials and Technology, South China Academy of Advanced Optoelectronics, South China Normal University, Guangzhou 510006, China*

[2]*Laboratory of Solid State Microstructures, Nanjing University, Nanjing 210093, China*

[3]*National Center for International Research on Green Optoelectronics, South China Normal University, Guangzhou 510006, China*

Email: qinmh@scnu.edu.cn

**[Abstract]** The controllable magnetic skyrmion motion represents a highly concerned issue in preparing advanced skyrmion-based spintronic devices. Specifically, magnon-driven skyrmion motion can be easily accessible in both metallic and insulating magnets, and thus is highly preferred over electric current control further for the ultra-low energy consumption. In this work, we investigate extensively the dynamics of skyrmion motion driven by magnon in an antiferromagnet using the collective coordinate theory, focusing on the effect of magnon polarization. It is revealed that the skyrmion Hall motion driven by circularly polarized magnon becomes inevitable generally, consistent with earlier report. Furthermore, the elastic scattering theory and numerical results unveil the strong inter-dependence between the linearly polarized magnon and skyrmion motion, suggesting the complicated dependence of the skyrmion motion on the polarization nature of driving magnon. On the reversal, the scattering from the moving skyrmion may lead to decomposition of the linearly polarized magnon into two elliptically polarized magnon bands. Consequently, a net transverse force acting on skyrmion is generated owing to the broken mirror symmetry, which in turn drives a skyrmion Hall motion. The Hall motion can be completely suppressed only in some specific condition where the mirror symmetry is preserved. The present work unveils non-trivial skyrmion-magnon scattering behavior in antiferromagnets, advancing the antiferromagnetic spintronics and benefiting to high-performance devices.

Keywords: antiferromagnetic skyrmion, magnon, spintronics

**I. Introduction**

During the past decades, the dynamics of skyrmion [1] has attracted extensive attention for designing advanced skyrmion-based spintronic devices such as race-track memory and logic units [2,3]. Specifically, ferromagnetic skyrmions have been observed in a series of chiral magnets [4-8] and heavy metal/ferromagnetic films [9-12] with broken inversion symmetry. Effective controls of skyrmions have been demonstrated using various external stimuli including electric current [13,14], gradient magnetic field [15], oscillating and gradient electric field [16,17], and polarized magnon [18-23]. Among these stimuli, magnons as the quanta of spin waves, driving skyrmion motion without Joule heating due to the absence of charge physical transport, are particularly attractive for the advantage of low-energy consumption.

The driving force of skyrmion motion exerted by magnons in ferromagnets can be understood by considering the deflection of polarized magnons by the fictitious magnetic field from the neighboring skyrmion, which on the contrary effectively drives the skyrmion motion through the momenta exchange [18,24,25]. Alternatively, antiferromagnetic (AFM) skyrmions have also been theoretically predicted [26,27] and experimentally observed in synthetic antiferromagnets [28], which are of great interest in high-density and high-speed [29,30] spintronic devices attributing to their particular merits such as strong anti-interference capability and ultrafast magnetic dynamics [31,32]. Therefore, an understanding of the skyrmion motion driven by the AFM magnons, no matter how they are generated, becomes highly concerned. Interestingly, an AFM skyrmion moves only longitudinally with electric current due to its two-sublattice spin structures which are of opposite topological numbers and experience well cancelled Magnus forces. Nevertheless, it should be mentioned that magnon-driving rather than electric current control would be highly preferred not only for energy saving, but also for that magnon control works much better in insulating systems, noting that plenty of antiferromagnets are insulating.

Unlike the case in ferromagnets where magnons can only be right-circularly polarized, magnons in antiferromagnets can be both right- and left-circularly polarized, adding a polarization degree of freedom including all linear and elliptical polarizations [33,34]. Thus, this degree of freedom could be used in modulating skyrmion dynamics and even in encoding information in magnons. For instance, in skyrmion lattice phases of antiferromagnets, left- and right-hand magnons are deflected toward opposite transverse directions due to their opposite effective charges, resulting in the so-called magnon spin Hall effect. Consequently, injected circularly polarized magnons can effectively drive the skyrmion Hall motion even in antiferromagnets. Furthermore, it is demonstrated that the Hall motion highly depends on magnon polarization, which could not be possible for linearly polarized magnons that attribute to the equal superposition of the left- and right-handed magnon bands, similar to the current driven AFM skyrmion motion [35].

These important works thus unveil interesting magnon-driven skyrmion dynamics, definitely benefiting to future spintronic and magnonic applications. However, the interplay between polarized magnons and AFM skyrmions could be more complex. Theoretically, linearly polarized magnons could be possibly decomposed into magnon bands with elliptical polarizations, which breaks the mirror symmetry and generates a net transverse force on the topological spin texture. In this case, different spin dynamics such as skyrmion Hall motion could be induced by linearly polarized magnons. As a matter of fact, magnon scattering by AFM domain wall has been revealed to be strongly dependent on the linear polarization of injected magnons [36,37]. In the presence of the Dzyaloshinskii-Moriya interaction (DMI), the magnon with the out-of-plane linear polarization is reflected by domain wall due to the increased scattering potential and drives the wall forward, while the in-plane linearly polarized magnons propagate through the wall almost freely [36]. In view of the relevance between these noncollinear magnetic textures, strong dependence of skyrmion dynamics on linear polarization direction of magnons is expected. Thus, the effect of magnon polarization on skyrmion dynamics urgently deserves to be clarified, considering its importance in AFM spintronics and magnonics.

In this work, we investigate the skyrmion motion driven by injected magnons in antiferromagnets using analytical methods and the Landau-Lifshitz-Gilbert (LLG) simulations, and pay particular attention onto the effect of magnon polarization. The skyrmion Hall motion equation driven by circularly polarized magnon is derived based on the collective coordinate theory. Moreover, the strong inter-dependence between the linearly polarized magnon and skyrmion longitudinal motion is theoretically and numerically revealed, allowing a comprehensive understanding of the driving mechanism. More interestingly, we numerically demonstrate that linearly polarized magnons can be generally decomposed into two elliptically polarized magnon bands with opposite handedness. As a result, a net transverse force acting on the skyrmion is induced owing to the broken mirror symmetry, which in turn drives the skyrmion Hall motion.

## II. Model and methods

Similar to the earlier work [38], we study a two-dimensional AFM model in the *xy*-plane with two magnetic sublattices that have magnetic moments $\mathbf{m}_1$ and $\mathbf{m}_2$ respectively, satisfying condition $|\mathbf{m}_1| = |\mathbf{m}_2| = S$ with spin length $S$. The normalized staggered Néel vector $\mathbf{n}$ is defined as $\mathbf{n} = (\mathbf{m}_1 - \mathbf{m}_2)/2S$ [39] to describe the Lagrangian. Taking into account the exchange energy, the anisotropy energy, and the interfacial DMI as well, one has the Lagrangian density $L$ given by [40]:

$$L = \frac{\rho_0^2}{2A_0}\dot{\mathbf{n}}^2 - u_0, \tag{1}$$

with the thermodynamic free energy:

$$u_0 = A^*(\nabla \mathbf{n})^2/2 - Kn_z^2/2 + D(n_z \nabla \cdot \mathbf{n} - (\mathbf{n} \cdot \nabla)n_z)/2, \tag{2}$$

where $A_0$ and $A^*$ are the homogeneous and effective exchange constants respectively, $K$ is the easy *z*-axis anisotropy constant, $D$ is the DMI constant, and $\rho_0 = \hbar S/a$ is the density of the staggered spin angular momentum per unit cell [41] with lattice constant $a$ and reduced Planck's constant $\hbar$.

To describe the magnons, it is convenient to use a global frame defined by three mutually orthogonal unit vectors ($\mathbf{e}_1$, $\mathbf{e}_2$, $\mathbf{e}_3$) with $\mathbf{e}_3 = \mathbf{n}_0/|\mathbf{n}_0| = \mathbf{e}_1 \times \mathbf{e}_2$ where $\mathbf{n}_0$ is the equilibrium configuration. A weakly excited state can be parametrized as $\mathbf{n} = \mathbf{n}_0 + \delta_x \mathbf{e}_1 + \delta_y \mathbf{e}_2$, where $\delta_x$ and $\delta_y$ describe the amplitude components of magnon. Then, one obtains the two monochromatic solutions with the complex fields: $\psi^* = \delta_x + i\delta_y$ for right-circularly polarized magnon and $\psi = \delta_x - i\delta_y$ for left-circularly polarized magnon [42]. Moreover, the complex field can also be rewritten in the form of plane wave $\psi = \exp[i(\mathbf{k}\cdot\mathbf{r} - \omega t)]$ with wave vector $\mathbf{k}$, position vector $\mathbf{r}$ with length $r$ and polar angle $\vartheta$, frequency $\omega$, and time $t$.

Subsequently, the skyrmion dynamics induced by polarized magnons are analytically calculated using the collective coordinate theory and elastic scattering theory. Moreover, the position and velocity of the skyrmion are also estimated based on the LLG simulations of the discrete model, in order to check the validity of theoretical analysis. The LLG simulation details are presented in Appendix A.

### III. Results and discussion

*A. Hall motion of skyrmion*

A scheme of collective coordinates for magnon and skyrmion in a Lagrangian frame will be used to formulate the skyrmion dynamics driven by circularly polarized magnons, noting that such a scheme was once proposed to study the skyrmion-magnon scattering in ferromagnets [24]. Following the earlier work, we transform the $z$-axis to the equilibrium configuration $\mathbf{n_0}$ using the rotation matrix $\mathbf{T}$ satisfying $\mathbf{n_0} = \mathbf{T}\mathbf{n'_0}$ with $\mathbf{n'_0} = \mathbf{e}_z$ where $\mathbf{e}_z$ is the unit vector along the $z$-axis [43-45], in order to conveniently derive the emergent electromagnetic field.

The Lagrangian density $L$ can be divided into three parts: $L_0 = (\rho_0 \partial_t \mathbf{n_0})^2/2A_0 - u_0$ from the equilibrium texture, $L_{sw} = \rho_0^2(\partial_t\psi^* \cdot \partial_t\psi)/2A_0 - A^*(\partial_i\psi^* \cdot \partial_i\psi)/2 + K(\psi^*\psi)/2$ from the disturbance part where $i = 1, 2$, and $3$ denote the spatial derivatives with respect to the $x$, $y$, and $z$-axes respectively, and $L_{int} = -i\rho_0^2(\psi^*\partial_t\psi - \psi\partial_t\psi^*)\,a_t^0/2A_0 - \mathbf{j}\cdot\mathbf{a}_{total} + \psi^*\psi u_0 - \psi^*\psi(\rho_0\partial_t\mathbf{n_0})^2/2A_0$ from the skyrmion-magnon interactions (the detailed derivation is presented in Appendix B). Here, $a_t^0 = -\cos\theta \cdot \partial_t\varphi$ [24,44,46,47] coincides with the geometrical scalar potential, due to the basis variation, $\theta$ and $\varphi$ are the polarization angle and azimuth angle of $\mathbf{n}$ respectively, $\mathbf{a}_{total} = \mathbf{a}^0 + \mathbf{a}_D$ [44] is the total vector potential including the contributions from the inhomogeneous magnetization $\mathbf{a}^0 = -\cos\theta \cdot \nabla\varphi$ and from the DMI $\mathbf{a}_D = -(D/2A^*)\mathbf{n_0}$, and $\mathbf{j} = -iA^*(\psi^*\nabla\psi - \psi\nabla\psi^*)/2$ is the spin wave flux.

For a stable skyrmion, the Thiele theory can be used to describe its dynamics, and its position is characterized by **X** = {$X_i$}. Similarly, one may use a set of collective coordinates **x** = {$x_u$} to characterize the position of a spin-wave packet [24] and to estimate the magnon group velocity $\partial\omega/\partial\mathbf{k}$. Then, the full Lagrangian $L_z$ can be rewritten in terms of {$X_i$} and {$x_u$},

$$L_z = M^{ij}\dot{X}_i^2 - 2\omega A_i^0 \dot{X}_i - U_0 + \rho[(\rho_0\omega)^2/A_0 - A^*k^2] - 2\rho A^* \mathbf{a}_{\text{total}} \cdot \mathbf{k}, \quad (3)$$

where the Einstein summation rules over the repeated indices, $M^{ij} = (1-2\rho)\rho_0^2 \int dV(\partial_i \mathbf{n}_0 \cdot \partial_j \mathbf{n}_0)/2A_0$ is the dissipative tensor describing the effective mass [48] due to the exchange interaction between neighboring spins, $\dot{X}$ represents the derivative with respect to time, $U_0 = \int dV(1-\psi\psi^*)u_0$ is the total texture energy, $A_i^0 = \rho_0^2 \int dV \psi\psi^* a_i^0/2$, $A_0$ is the vector potential, and $\rho = \int dV \psi\psi^*/2$ is the local intensity.

Similarly, the Rayleigh function is rewritten as

$$R = \frac{\alpha}{2}[\frac{2A_0}{\rho_0^2}M^{ij}\dot{X}_i\dot{X}_j + (\frac{\omega^2}{2A^*})\psi^*\psi\dot{x}_u^2], \quad (4)$$

After applying the Euler-Lagrange rule, we obtain the two coupled equations to describe the dynamics of skyrmion and magnon:

$$2M^{ij}\ddot{X}_j + \partial_i U_0 + 2\omega\left(\partial A_j^0/\partial X_i - \partial A_i^0/\partial X_j\right)\dot{X}_j + 2\alpha A_0 M^{ij}\dot{X}_j/\rho$$
$$= -2\rho A^* m_{\text{sw}}\left(\partial a_v/\partial X_i - \partial a_i/\partial x_v\right)\dot{x}_v \quad (5a)$$

$$\frac{4\rho_0^4\omega^2}{A_0^2 A^*}\ddot{x}_u - 2\partial_u u_0 - 2\frac{\rho_0^2\omega}{A_0}\left(\partial a^0_v/\partial x_u - \partial a^0_u/\partial x_v\right)\dot{x}_v - \frac{\alpha\omega^2}{A^*}\dot{x}_u$$
$$= -2\omega\left(\partial a_u/\partial X_j - \partial a_j/\partial x_u\right)\dot{X}_j \quad (5b)$$

where $m_{\text{sw}}$ is the effective mass of the magnon.

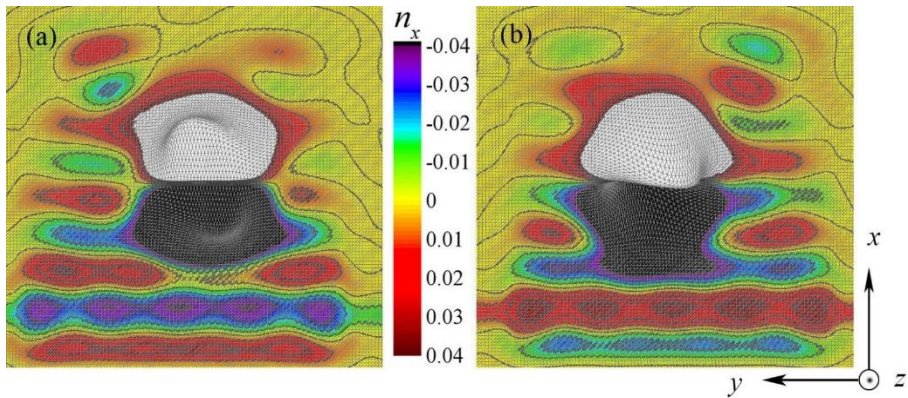

Fig. 1. The spatial map of $n_x$ for the left-circularly (a) and right-circulary (b) polarized magnons injected from the bottom of the frame and scattered by the AFM skyrmion (the black and white part).

In some extent, the motion equation is equivalent to the classical motion of a massive particle, subject to dissipation-induced friction and external forces. Particularly, the third term in the left side of Eq. (5b) is associated with the effective Lorentz force acting on the spin wave packet caused by the effective magnetic field of the skyrmion, which induces a transverse motion of the packet. Moreover, the effective field is reversed when the sign of $\omega$ is changed, resulting in the topological spin Hall effect as depicted in Fig. 1 where the LLG simulated spatial map of $n_x$ is presented. It is clearly shown that the left-circularly polarized magnons with $\omega > 0$ are deflected to the left side (Fig. 1(a)) by the skyrmion, while the right-circularly polarized magnons with $\omega < 0$ are deflected to the right side (Fig. 1(b)), consistent with the earlier report [35].

For an injected magnon current, Eq. (5a) is updated to:

$$2M^{ij}\ddot{\mathbf{X}} + \partial_i U_0 + \dot{\mathbf{X}} \times 2\omega\rho(4\pi Q \cdot \mathbf{e}_z) + 2\alpha A_0 M^{ij}\dot{\mathbf{X}}/\rho = -2\rho A^* m_{sw}\dot{\mathbf{x}} \times (4\pi Q \cdot \mathbf{e}_z), \qquad (6)$$

where $\ddot{\mathbf{X}}$ is the second-order derivative with respect to time, and $Q = \int dV(\nabla \times \mathbf{a}^0)_z/4\pi$ is the staggered topological charge. It is noted that the right term is the driving force acting on the skyrmion from the injected magnons, which can be divided into longitudinal and transverse parts. Thus, the Magnus force (the third term in the left side) is reversed for opposite $\omega$ [35,49] or opposite $Q$, so does the transverse driving force due to the topological spin Hall effect. As a result, a skyrmion Hall motion is induced by the injected magnons, which can be tuned by the magnon handedness and skyrmion charge.

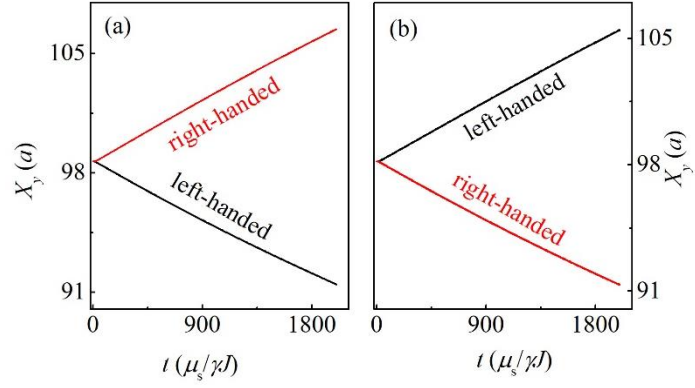

Fig. 2. The temporal evolution of the position $X_y$ of (a) skyrmion, and (b) anti-skyrmion for the left-handed and right-handed polarized magnons.

The dependences of the skyrmion Hall motion on $\omega$ and $Q$ are verified by the LLG simulations. Fig. 2(a) gives the transverse position of the skyrmion, demonstrating the opposite transverse motions respectively driven by the left-handed magnons and right-handed magnons. On the other hand, for an anti-skyrmion with an opposite $Q$ [50], the transverse motion driven by the magnons is opposite to that of the skyrmion, as shown in Fig. 2(b).

Considering a stable skyrmion motion and neglecting $\partial_i U_0$, the solution to Eq. 6 gives the skyrmion velocity:

$$\mathbf{v} = \frac{\rho \mathbf{F}_m}{\alpha M^{ij}}, \tag{7}$$

where $\rho$ is linearly related to $h^2$ with magnetic field $h$ used to generate magnons [51], and $\mathbf{F}_m = 8\pi A^* m_{sw} \dot{\mathbf{x}} \times (Q \cdot \mathbf{e}_z) + 8\pi \omega \mathbf{v} \times (Q \cdot \mathbf{e}_z)$ is the driving force. Thus, a linear dependence of the skyrmion speed $v$ on $h^2/\alpha$ is expected in case of weak field $h$.

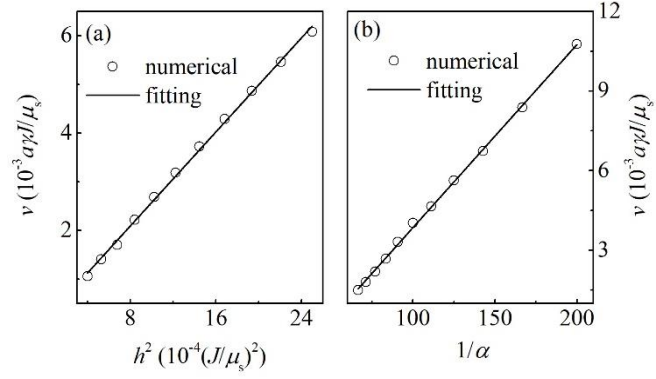

Fig. 3. The simulated (empty circles) and fitted (solid line) skyrmion speeds $v$ as functions of (a) $h^2$ for $\alpha = 0.01$, and (b) $1/\alpha$ for $h = 0.04J/\mu_s$.

In order to check the validity of theory, the dependences of $v$ on various parameters are numerically simulated and analytically fitted. Fig. 3(a) gives the simulated $v$ for various $h^2$, demonstrating a linear relation. It is noted that the magnon amplitude is linearly correlated with $h$, and the driving force on skyrmion is related to $h^2$, resulting in the linear dependence of $v$ on $h^2$. Moreover, the skyrmion mobility is reduced by an enhanced damping term, resulting in the decrease of $v$ with the increasing $\alpha$, as shown in Fig. 3(b) which demonstrates a linear relation between $v$ and $1/\alpha$.

## B. Longitudinal motion of skyrmion driven by linearly polarized magnon

In this section, we analytically investigate the skyrmion longitudinal motion driven by linearly polarized magnons. Here, the remaining massive fluctuation modes are represented by the dimensionless complex field $\psi$. The Néel vector is expressed as:

$$n_x = (2\sin^2\frac{\theta}{2}\cos^2\varphi - 1)\delta_x + 2\sin^2\frac{\theta}{2}\cos\varphi\sin\varphi\delta_y + \sin\theta\cos\varphi\delta_z$$
$$n_y = 2\sin^2\frac{\theta}{2}\cos\varphi\sin\varphi\delta_x + (2\sin^2\frac{\theta}{2}\sin^2\varphi - 1)\delta_y + \sin\theta\sin\varphi\delta_z , \qquad (8)$$
$$n_z = \sin\theta\cos\varphi\delta_x + \sin\theta\sin\varphi\delta_y + \cos\theta\delta_z$$

with $\delta_z = (1 - \delta_x^2 - \delta_y^2)^{1/2}$.

First, we study the *x*- and *y*-linearly polarized magnons driven skyrmion motion. For an elastic scattering, substituting Eq. (8) into Eq. (1) and conserving the second order in the fluctuation field, one obtains the Hamiltonian densities for the *x*- and *y*-linear polarizations $L_x$ and $L_y$ respectively:

$$\begin{aligned}
L_x = &-A^*\{-\sin^2\varphi(\partial_r\theta)^2 + [2\cos\theta(\cos\theta-1) + \sin^2\theta\cos^2\varphi]/r^2\}\delta_x^2/2 \\
&+ \rho_0^2\dot{\delta}_x^2/2A_0 - D[-\sin^2\varphi\partial_r\theta + \sin\theta(1-\cos\theta-\cos\theta\cos^2\varphi)/r]\delta_x^2/2 \\
&+ A^*\delta_x\nabla^2\delta_x/2 + K(\sin^2\theta\cos^2\varphi - \cos^2\theta)\delta_x^2/2 \\
L_y = &-A^*\{-\cos^2\varphi(\partial_r\theta)^2 + [2\cos\theta(\cos\theta-1) + \sin^2\theta\sin^2\varphi]/r^2\}\delta_y^2/2 \\
&+ \rho_0^2\dot{\delta}_y^2/2A_0 - D[-\cos^2\varphi\partial_r\theta + \sin\theta(1-\cos\theta-\cos\theta\sin^2\varphi)/r]\delta_y^2/2 \\
&+ A^*\delta_y\nabla^2\delta_y/2 + K(\sin^2\theta\sin^2\varphi - \cos^2\theta)\delta_y^2/2
\end{aligned} \quad (9)$$

Applying the Euler-Lagrangian equation, we obtain the dynamic equations for the *x*- and *y*-linearly polarized magnons respectively:

$$\begin{aligned}
\rho_0^2\ddot{\delta}_x/A_0 = &-A^*\{-\sin^2\varphi(\partial_r\theta)^2 + [2\cos\theta(\cos\theta-1) + \sin^2\theta\cos^2\varphi]/r^2\}\delta_x \\
&- D[-\sin^2\varphi\partial_r\theta + \sin\theta(1-\cos\theta-\cos\theta\cos^2\varphi)/r]\delta_x \\
&+ A^*\nabla^2\delta_x + K(\sin^2\theta\cos^2\varphi - \cos^2\theta)\delta_x \\
\rho_0^2\ddot{\delta}_y/A_0 = &-A^*\{-\cos^2\varphi(\partial_r\theta)^2 + [2\cos\theta(\cos\theta-1) + \sin^2\theta\sin^2\varphi]/r^2\}\delta_y \\
&- D[-\cos^2\varphi\partial_r\theta + \sin\theta(1-\cos\theta-\cos\theta\sin^2\varphi)/r]\delta_y \\
&+ A^*\nabla^2\delta_y + K(\sin^2\theta\sin^2\varphi - \cos^2\theta)\delta_y
\end{aligned} \quad (10)$$

For a monochromatic magnon, solving the dynamic equation becomes computing the eigen-problem of $H\delta_{x,y} = \rho_0^2\omega^2\delta_{x,y}/A_0$ with $H = H_0 + H_{sx,sy}$ where $H_0 = A^*/r^2 - A^*\nabla^2 + K$ is the ground-state Hamiltonian, and the skyrmion scattering potentials for the *x*- and *y*-linear polarizations $H_{sx}$ and $H_{sy}$ read respectively:

$$\begin{aligned}
H_{sx} = &A^*\{-\sin^2\varphi(\partial_r\theta)^2 + [2\cos\theta(\cos\theta-1) + \sin^2\theta\cos^2\varphi+1]/r^2\} \\
&+ D[-\sin^2\varphi\partial_r\theta + \sin\theta(1-\cos\theta-\cos\theta\cos^2\varphi)/r] \\
&- K(\sin^2\theta\cos^2\varphi - \cos^2\theta+1) \\
H_{sy} = &A^*\{-\cos^2\varphi(\partial_r\theta)^2 + [2\cos\theta(\cos\theta-1) + \sin^2\theta\sin^2\varphi+1]/r^2\} \\
&+ D[-\cos^2\varphi\partial_r\theta + \sin\theta(1-\cos\theta-\cos\theta\sin^2\varphi)/r] \\
&- K(\sin^2\theta\sin^2\varphi - \cos^2\theta+1)
\end{aligned} \quad (11)$$

Thus, different scattering behaviors and skyrmion longitudinal speeds are expected due to the different scattering potentials for the *x*- and *y*-linear polarizations.

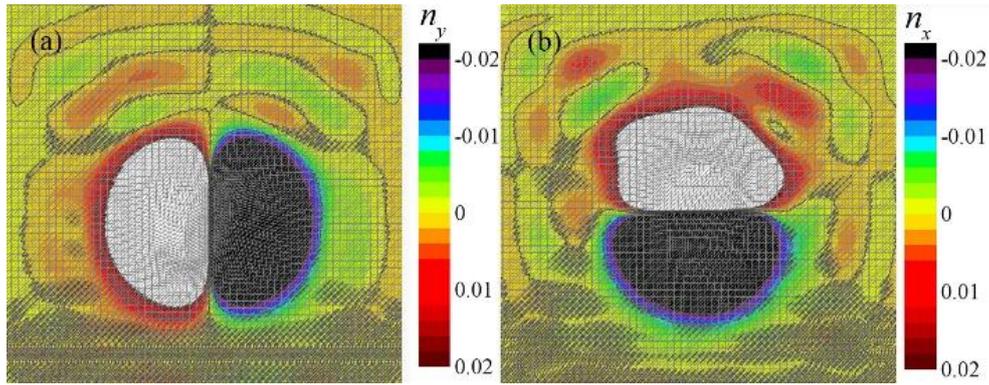

Fig. 4. Spatial map of (a) $n_y$ for the *x*-linearly, and (b) $n_x$ for the *y*-linearly polarized magnons scattered by the skyrmion.

It is worth noting that the elastic scattering with an unchanged magnon polarization is simply considered in the derivation of the Lagrangian density, which is not exactly consistent with the fact that the linearly polarized magnon is generally decomposed into circularly or elliptically polarized magnon bands by the skyrmion, due to the topological spin Hall effect. However, the analytical argument on the skyrmion longitudinal motion works qualitatively at least in such inelastic scattering process. For example, different scattered magnon amplitudes for the *x*- and *y*-linear polarizations are observed in the LLG simulations, as demonstrated in Fig. 4 where the spatial evolution of the components of **n** is plotted. It is clearly shown that the linearly polarized magnons are decomposed into the left- and right-handed magnon bands, while the scattered magnon amplitude for the *y*-linear polarization (Fig. 4(b)) is obviously larger than that for the *y*-linear polarization (Fig. 4(a)).

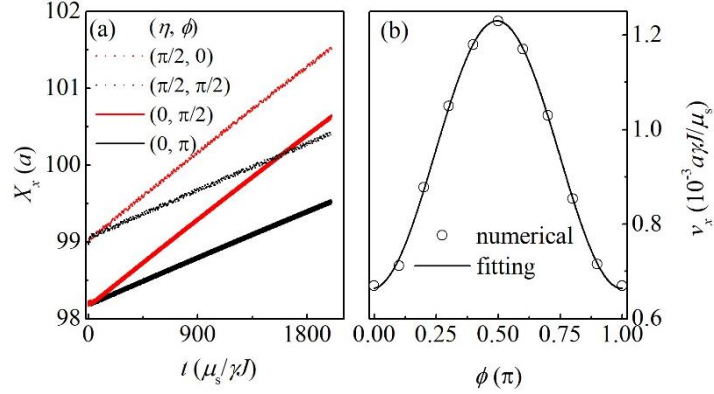

Fig. 5. (a) The Position $X_x$ of the Néel skyrmion (solid lines) and Bloch skyrmion (dotted lines) as functions of time driven by the *x*- and *y*-linearly polarized magnons. (b) The Simulated (empty circles) and analytically fitted (solid line) $v_x$ as functions of $\phi$.

Consequently, the longitudinal speed of the skyrmion is also dependent of the linear polarization direction of the injected magnons, noting that the driving force is related to the scattered magnon amplitude. In Fig. 5(a), the simulated skyrmion position (solid lines with the Néel skyrmion helicity $\eta = 0$, and $\phi$ is the angle between the polarization direction and *x* axis) driven by the *x*- and *y*-linearly polarized magnons are presented, demonstrating that the skyrmion motion for the *y*-linear polarization with $\phi = \pi/2$ is much faster than the *x*-linear polarization with $\phi = \pi$, consistent with the magnon scattering behaviors.

Generally, the skyrmion scattering potential depending on the polarization direction reads:

$$H_s = A^*\{[\sin^2(\vartheta+\phi+\eta)-1](\partial_r\theta)^2 \\ +[2\cos\theta(\cos\theta-1)+\sin^2\theta\sin^2(\vartheta+\phi+\eta)-\sin^2\theta+1]/r^2\} \\ + D[[\sin^2(\vartheta+\phi+\eta)-1]\partial_r\theta + \sin\theta(1-\cos\theta-\cos\theta\sin^2(\vartheta+\phi+\eta))/r] \\ - K(\sin^2\theta\sin^2(\vartheta+\phi+\eta)-\cos^2\theta+1)$$
(12)

Thus, $H_s$ is dependent on the skyrmion helicity $\eta$ and magnon polarization direction, and equal $\eta + \phi$ will result in equal skyrmion longitudinal speeds $v_x$, which has been confirmed in the LLG simulated dynamics of Bloch skyrmion motion with $\eta = \pi/2$ shown in Fig. 5(a) (dotted lines). It is clearly demonstrated that the longitudinal speed of the Bloch skyrmion driven by the $x$-/$y$-linearly polarized magnons is the same as that of the Néel skyrmion driven by the $y$-/$x$-linearly polarized magnons, confirming above theoretical argument.

As a matter of fact, $v_x$ depending on $\phi$ could be reasonably assumed to be $v_x = C_1 \sin^2(\vartheta + \eta + \phi_1) + C_2$ with the parameters $C_1$ and $C_2$ independent of $\phi$, considering the fact that the momentum transfer between skyrmion and magnons is mainly determined by the scattering potential. The LLG simulated $v_x$ for various $\phi$ are summarized in Fig. 5(b), which can be well fitted by this equation, further confirming the validity of the elastic theory in studying the skyrmion longitudinal speed driven by linearly polarized magnons.

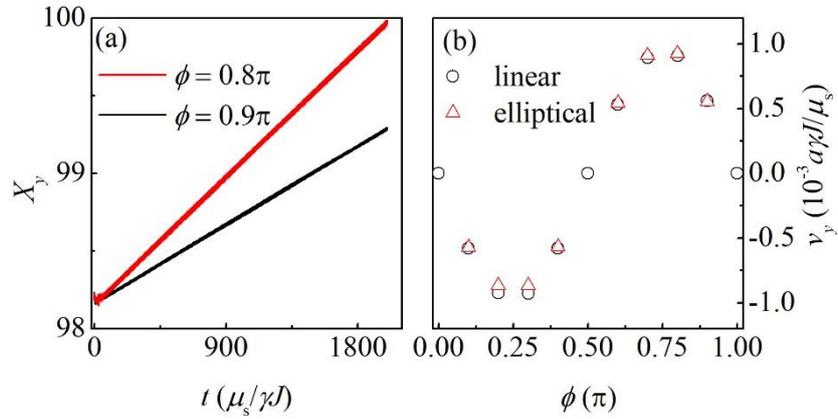

Fig. 6. (a) The evolutions of $X_y$ for $\phi = 0.8\pi$ and $0.9\pi$, and (b) The simulated $v_y$ as functions of $\phi$ driven by the linearly polarized magnons (black empty circles) and by isolated elliptically polarized magnons (red empty triangles).

## C. Transverse motion of skyrmion driven by linearly polarized magnon

Interestingly, a skyrmion transverse motion driven by linearly polarized magnons with a polarization deviated from the $x$-axis and $y$-axis is generally observed, as shown in Fig. 6(a) which gives the LLG simulated evolutions of the skyrmion position $X_y$ for $\phi = 0.8\pi$ and $0.9\pi$. The skyrmion transverse motion is non-negligible and strongly dependent on the polarization direction. In Fig. 6(b), the simulated transverse speed $v_y$ as a function of $\phi$ is presented, demonstrating a sinusoidal dependence of $v_y$ on $2\phi$. This behavior is somewhat contrary to the earlier viewpoint that the AFM skyrmion may have no transverse motion when the injected magnons are linearly polarized.

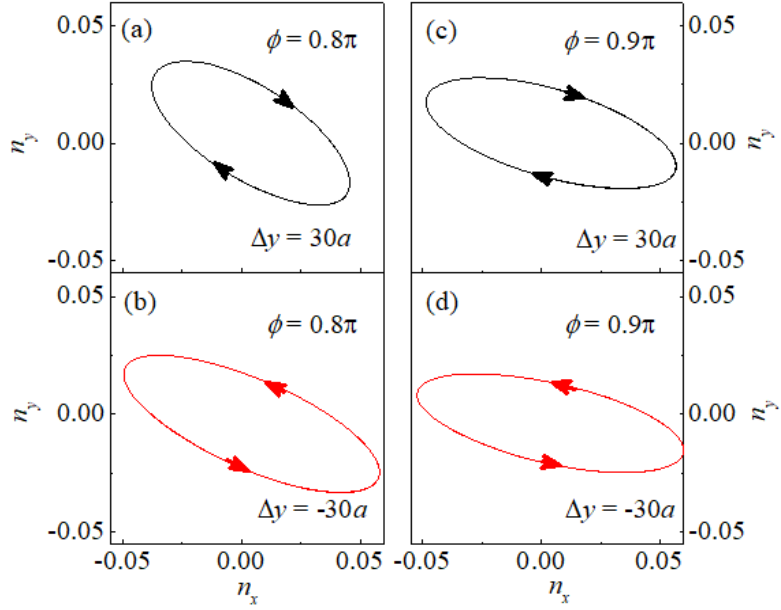

Fig. 7. The local magnetization-precession trajectory at different positions for $\phi = 0.8\pi$ ((a) and (b)) and $\phi = 0.9\pi$ ((c) and (d)).

In order to understand this unexpected behavior, we trace the local magnetization-precession trajectories at different positions, and give the results in Fig. 7. Unlike the former view that linearly polarized magnons are generally decomposed by AFM skyrmion into circularly polarized magnon bands, the simulations clearly demonstrate that the injected magnons are decomposed into elliptically polarized magnon bands with opposite handedness, as shown in Figs. 7(a) and 7(b) where give the magnetization-trajectories at $x$-axis symmetric positions for $\phi = 0.8\pi$, respectively.

In this case, the left-handed/right-handed magnons scattered by the skyrmion can be described by $\psi_{r/l} = A_1 \cos(\mathbf{k}\cdot\mathbf{x} - \omega t \mp \varphi_0)\mathbf{n}_x' + A_2 \cos(\mathbf{k}\cdot\mathbf{x} - \omega t)\mathbf{n}_y'$ with the amplitudes $A_1$ and $A_2$, and the phase $\varphi_0$ are related to $\phi$, as demonstrated in Figs. 7(c) and 7(d) where give the local magnetization-precession trajectories for $\phi = 0.9\pi$. Thus, the decomposed magnon bands break mirror symmetry and in turn generate a net transverse force acting on the skyrmion, resulting in a skyrmion transverse motion, as revealed in our simulations. As a matter of fact, the transverse motion driven by two isolated magnon currents with left- and right-hand elliptical polarizations is also simulated, and the results are given in Fig. 6(b). The well consistence between the simulated $v_y$ for the linear polarization and elliptical polarization strongly supports the newly revealed skyrmion-magnon scattering picture.

Particularly, for the *x*- and *y*-linearly polarized magnons, the decomposed spin waves are parameterized by $\psi_{r/l} = A\cos(\boldsymbol{k}.\boldsymbol{x} - \omega t \mp \pi/2)\mathbf{n}_x' + A\cos(\boldsymbol{k}.\boldsymbol{x} - \omega t)\mathbf{n}_y'$ and $\psi_{r/l} = A\cos(\boldsymbol{k}.\boldsymbol{x} - \omega t)\mathbf{n}_x' + A\cos(\boldsymbol{k}.\boldsymbol{x} - \omega t \mp \pi/2)\mathbf{n}_y'$, respectively, which are exactly corresponding to circularly polarized magnons. In each case, mirror symmetry is preserved, and the transverse forces from the left-handed and right-handed magnon bands acting on the skyrmion are perfectly canceled out, resulting in the absence of skyrmion Hall motion, as have been numerically revealed in the earlier work.

Indeed, skyrmion Hall effect may prohibit a precise control of skyrmion motion, which goes against future applications. Interestingly, electric current drives an AFM skyrmion motion straight along the current direction without path deviation, making AFM skyrmion attractive to device design. However, electric current control is deficient for energy loss and only works in metallic antiferromagnets, and thus other energy-saving controls such as using polarized magnon are highly preferred. Importantly, the current simulations suggest that linear polarization of injected magnons should be delicately tuned to completely diminish skyrmion Hall effects. Thus, this work further clarifies the complex interplay between skyrmion and polarized magnon in antiferromagnets, which is very meaningful for spintronic and magnonic applications.

## IV. Conclusion

In conclusion, we have studied the skyrmion dynamics induced by polarized magnons in antiferromagnets using analytical methods and numerical simulations. The skyrmion Hall motion driven by circularly polarized magnons are explained based on the collective coordinate theory. In addition, the skyrmion longitudinal motion strongly depending on the linear polarization of the injected magnons is revealed analytically and numerically. More interestingly, we demonstrate that the linearly polarized magnons generally drive a skyrmion Hall motion which is also dependent on the polarization direction. Thus, the present work unveils a new skyrmion-magnon scattering mechanism in antiferromagnets, benefiting to future spintronic applications.

## Acknowledgment

We sincerely appreciate the insightful discussions with Huaiyang Yuan, Zhengren Yan and Weihao Li. The work is supported by the Natural Science Foundation of China (Grants No. 51971096 and No. 51721001), and the Natural Science Foundation of Guangdong Province (Grant No. 2019A1515011028), and the Science and Technology Planning Project of Guangzhou in China (Grant No. 201904010019), and Special Funds for the Cultivation of Guangdong College Students Scientific and Technological Innovation (Grant No. pdjh2020a0148).

**Appendix A: Numerical simulations of the atomistic spin model**

In order to check the validity of the theory, we also perform the numerical simulations of the discrete model. Here, the two dimensional Hamiltonian of the atomistic spin model is given by

$$H = J\sum_{\langle i,j \rangle} \mathbf{S}_i \cdot \mathbf{S}_j - D_0 \sum_i (\mathbf{S}_i \times \mathbf{S}_{i+x} \cdot \mathbf{e}_y - \mathbf{S}_i \times \mathbf{S}_{i+y} \cdot \mathbf{e}_x) - K_0 (S_i^z)^2, \tag{A1}$$

where the first term is the exchange interaction with $J = 1$ between the nearest neighbor spins, the second term is the DMI with $D_0 = 0.11\ J$ and the unit vector $\mathbf{e}_{x/y}$ along the $x/y$ axis, and the last term is the anisotropy energy with $K_0 = 0.02\ J$. The dynamics of the AFM skyrmion is investigated by solving the LLG equation,

$$\frac{\partial \mathbf{S}_i}{\partial t} = -\gamma \mathbf{S}_i \times \mathbf{H}_i + \alpha \mathbf{S}_i \times \frac{\partial \mathbf{S}_i}{\partial t}, \tag{A2}$$

where $\mathbf{H}_i = -\mu_s^{-1} \partial H/\partial \mathbf{S}_i$ is the effective field, and $\alpha = 0.01$ is the damping constant. We use the fourth-order Runge-Kutta method to solve the LLG equation on a 200 × 200 square lattice.

The magnons are excited by a homogeneous and dimensionless magnetic field source. Specifically, we generate right-/left-handed magnons by applying ac magnetic field $\mathbf{h}_R/\mathbf{h}_L = h[\cos(\omega_0 t)\mathbf{e}_x \pm \sin(\omega_0 t)\mathbf{e}_y]$ with the frequency $\omega_0$. Similarly, the $y$- and $x$-linearly polarized magnons are generated by applying $\mathbf{h}_x = h\cos(\omega_0 t)\mathbf{e}_x$ and $\mathbf{h}_y = h\sin(\omega_0 t)\mathbf{e}_y$ respectively. The absorbing boundary conditions are used to eliminate the reflection of the magnons at boundary. The position of the skyrmion $\mathbf{X}$ is estimated by[20]

$$X_i = \frac{\int [i\mathbf{n} \cdot (\partial_x \mathbf{n} \times \partial_y \mathbf{n})]dxdy}{\int [\mathbf{n} \cdot (\partial_x \mathbf{n} \times \partial_y \mathbf{n})]dxdy}, \quad i = x, y \tag{A3}$$

Then, the velocity is numerically calculated by $\mathbf{v} = d\mathbf{X}/dt$.

**Appendix B: The derivation of the Lagrangian density**

In the local coordinate system, the Néel vector $\mathbf{n}$ reads $\mathbf{n}' = (\delta_x, \delta_y, \delta_z)$ with $\delta_z = (1 - \delta_x^2 - \delta_y^2)^{1/2}$. Here, the rotation matrix component is given by $\mathbf{T}_{ij} = 2P_i P_j - \delta_{ij}$ with $i = 1, 2, 3$ and the vector [23,44]

$$\mathbf{P} = (\sin\frac{\theta}{2}\cos\varphi, \sin\frac{\theta}{2}\sin\varphi, \cos\frac{\theta}{2}), \tag{B1}$$

and the Dirac delta function $\delta_{ij}$. The berry phase term is given by

$$L_b = \rho_0^2 \dot{\mathbf{n}}^2 / 2A_0 = \rho_0^2 [(\partial_t + i\mathbf{A})\mathbf{n}']^2 / 2A_0, \tag{B2}$$

where $\mathbf{A} = \mathbf{T}^{-1}\partial_t \mathbf{T}$ is the vector potential. Substituting $\mathbf{n}'$ into Eq. (B2) and conserving the second order in the fluctuation field, one obtains the berry phase

$$L_b = \rho_0^2[(1-\psi^*\psi)\dot{\mathbf{n}}_0^2 + i(\psi^*\partial_t\psi - \psi\partial_t\psi^*)\cos\theta\partial_t\varphi + \partial_t\psi^*\partial_t\psi]/2A_0. \tag{B3}$$

and the Hamiltonian density is [24]

$$\begin{aligned}u_{total} &= u_0 + iA^*(\psi^*\nabla\psi - \psi\nabla\psi^*)a_{total}/2 \\ &\quad -\psi^*\psi u_0 + A^*\partial_i\psi^*\partial_i\psi/2 - K\psi^*\psi/2\end{aligned}. \tag{B4}$$